\documentclass{aastex}          
\usepackage{spr-astr-addons}    

\begin{document}

\title{On the supernova remnants with flat radio spectra}

\shorttitle{On the SNRs with flat radio spectra}
\shortauthors{D. Oni\'c}

\author{D. Oni\'c\altaffilmark{}}

\altaffiltext{1}{Department of Astronomy, Faculty of Mathematics, University of Belgrade, Serbia}

\begin{abstract}
A considerable fraction of Galactic supernova remnants (SNRs) characterize flat spectral indices ($\alpha<0.5$). 
There are several explanations of the flat radio spectra of SNRs in the present literature. The most of models involve a 
significant contribution of the second-order Fermi mechanism but some of them also discuss high compressions ($>4$), 
contribution of secondary electrons left over from the decay of charged pions, as well as the possibility of thermal 
contamination. In the case of expansion in high density environment, intrinsic thermal bremsstrahlung could 
theoretically shape the radio spectrum of an SNR and also account for observable curved -- "concave up" radio spectra 
of some Galactic SNRs. This model could also shed a light on the question of flat spectral indices determined in some 
Galactic SNRs. On the other hand, present knowledge of the radio continuum spectra (integrated flux densities at different 
frequencies) of SNRs prevent definite conclusions about the significance of proposed models so the question on flat spectral 
indices still remains open. New observations, especially at high radio continuum frequencies, are expected to solve 
these questions in the near future. Finally, as there is a significant connection between the majority of Galactic SNRs 
with flat integrated radio spectrum and their detection in $\gamma$-rays as well as detection of radiative recombination 
continua in their X-ray spectra, the analysis of high energy properties of these SNRs is very important.
\end{abstract}

\keywords{ISM: supernova remnants -- radio continuum: general -- radio continuum: ISM}

\section{Introduction}

The spectra of SNRs in radio continuum are usually represented by a power-law, reflecting the pure 
synchrotron radiation from the SNR shell:\begin{equation}
S_{\nu}\propto\nu^{-\alpha},
\end{equation} where $S_{\nu}$ is the spatially-integrated flux density and $\alpha$ is the radio spectral index.

Test-particle diffusive shock acceleration (DSA) theory predicts that for strong shocks, radio spectral index $\alpha$ 
is approximately 0.5 (Bell 1978ab). Indeed, mean value of the observed radio spectral indices, for Galactic SNRs, is 
around 0.5 and correspondent distribution is roughly Gaussian \citep{b7}. On the other hand, the dispersion in that 
distribution is non-negligible (see Fig.\@ 1 in Reynolds et al.\@ 2012).

In the case of shocks with sufficiently low Mach number (less than around ten), steeper spectra 
($\alpha>0.5$) are generally expected. In test-particle DSA, for the synchrotron spectral index $\alpha$, the following relations hold:
\begin{equation}
\alpha=\frac{s-1}{2},\ \ s=\frac{\chi+2}{\chi-1},\ \ \chi=\frac{\gamma+1}{\gamma-1+\frac{2}{M^{2}}}\approx\frac{\gamma+1}{\gamma-1}, 
\end{equation}
where $\gamma$ is the post-shock thermal gas adiabatic index, $\chi$ is the compression ratio for parallel shock, $M$ 
is the (upstream adiabatic) Mach number that represents the ratio of shock velocity to the local sound speed and $s$ is the energy spectral 
index. On the other hand, only few old SNRs would be expected to have such weak shocks \citep{b16}. In fact, SNRs with steep spectrum 
($\alpha>0.5$) are usually young objects \citep{b7,b33,b34}. Steeper radio spectra may be explained by a cumulative effect of 
non-linear and oblique-shock steepening in the case of young SNRs \citep{b28}. An important prediction of the so called 
non-linear DSA (NDSA), where the reaction effects of cosmic-ray particle pressure is taken into account, is that the spectrum 
of young SNRs should flatten at higher energies so that a "concave up" spectrum is formed \citep{b15,b8}.

There is also a considerable number of Galactic SNRs with $\alpha<0.5$ \citep{b16}. Contamination with flat spectrum 
thermal emission can not be completely ruled out in all of the cases so it may be partially responsible for these lower 
(flat) spectral index values.

\begin{table*}
  \centering
  \caption{Properties of several Galactic SNRs with $\alpha<0.5$ that expand in high density environment: type, spectral 
index $\alpha$, low-frequency turn-over in overall spectrum (t.o.), distance $d$, mean diameter $D$, approximate age $t$, 
association/interaction with molecular cloud (MC), and detection in $\gamma$ rays ($\gamma$).}
  \begin{tabular}{@{}cccccccccc@{}}
  \hline
SNR&Type&$\alpha$&$\rm{t.o.}$&$d\ [\rm{kpc}]$&$D\ [\rm{pc}]$&$t\ [\rm{kyr}]$&MC&$\gamma$&Ref.\\
\hline
G6.4-0.1 (W28)&M-M&0.35$^{*}$&?&1.9&26.5&33-150&Y&Y&1--5\\
G18.8+0.3 (Kes 67)&S&0.40&N&14&57&16-100&Y&?&6--8\\
G31.9+0.0 (3C391)&M-M&0.49$^{\dagger}$&Y&8&$\sim$14&4&Y&Y&9--12\\
G34.7-0.4 (W44)&M-M&0.37&N&2.9&$\sim$26&20&Y&Y&13--16\\
G39.2-0.3 (3C396)&C&0.34&Y&6.2&$\sim$13&3&Y&?&13, 17--19\\
G43.3-0.2 (W49B)&M-M&0.46&Y&8-11&9-13&2.3&Y&Y&13, 20--22\\
G89.0+4.7 (HB21)&M-M&0.36&N&0.8&$\sim$25&19&Y&Y&23--25\\
G94.0+1.0 (3C434.1)&S&0.45&N&4.5&$\sim$39&25&Y?&?&13, 26--27\\
G189.1+3.0 (IC443)&M-M&0.38&Y&1.5&20&4&Y&Y&23, 28--31\\
\hline
\end{tabular}

\bigskip

\begin{flushleft}

{\small

(1) Abdo at al.~(2010a), (2) Dubner et al.~(2000), (3) Gabici et al.~(2010), (4) Giuliani et al.~(2010), (5) Sawada \& Koyama (2012), 
(6) Tian et al.~(2007), (7) Paron et al.~(2012), (8) Dubner et al.~(1999), (9) Brogan et al.\@ (2005), (10) Chen et al.~(2004), 
(11) Castro \& Slane (2010), (12) Green (2009), (13) Sun et al.~(2011), (14) Abdo et al.~(2010b), (15) Castelletti et al.~(2007), 
(16) Uchida et al.~(2012), (17) Su et al.~(2011), (18) Patnaik et al.~(1990), (19) Anderson \& Rudnick (1993), 
(20) Abdo at al.~(2010c), (21) Moffett \& Reynolds (1994), (22) Zhou et al.~(2011), (23) Gao et al.~(2011), 
(24) Reichardt et al.~(2012), (25) Pannuti et al.~(2010), (26) Foster (2005), (27) Jeong et al.~(2012), (28) Troja et al.~(2006), 
(29) Troja et al.~(2008), (30) Castelletti et al.~(2011), (31) Abdo et al.~(2010d)

\medskip

$*$ \citet{b7} noted that spectral index of this SNR is varying.\\
$\dagger$ Or 0.54 at frequencies higher and 0.02 at frequencies lower than 1 GHz (Sun et al.~2011).\\}

\end{flushleft}

\end{table*}

The most of the (shell, composite and mixed-morphology) SNRs with $\alpha<0.5$ are in fact (evolutionary) old objects 
expanding in a high density environment. Because the shock velocities are much lower than in young SNRs, the oblique-shock 
effects considered by Bell et al.\@ (2011) are unlikely to contribute to the flattening of the radio spectral indices of 
these SNRs. Also, in the case of older SNRs, it is not yet clear whether the mechanism for producing relativistic electrons 
is local acceleration at a shock front or compression of the existing population of Galactic background relativistic 
electrons (Leahy 2006 and references therein).

It must be noted that the majority of members of the so called mixed-morphology class of SNRs have flat spectral indices 
(see Table 4 in Vink 2012). The mixed-morphology (thermal composite) SNRs are characterized with bright (thermal) interiors 
in X-rays, and bright (non-thermal) shell-like radio morphology \citep{b18, b26}. They are mainly evolutionary old SNRs 
expanding in a high density environment (many of them interacting with molecular clouds) and many of them are detected in 
$\gamma$-rays \citep{b26}. In Greens's catalog of Galactic SNRs \citep{b7} these remnants are classified either as shell 
or composite type (they are not separated from the rest of SNRs).

It should be pointed out that, from the list of the SNRs with flat spectrum, plerions (pulsar wind or synchrotron nebulae) 
are excluded since they are not of interest for this paper. In fact, as the radio and X-ray emission from these objects 
are powered by the pulsar wind, not by the supernova explosion, they should not be referred as SNRs (at least in a classical 
context). For pulsar wind nebulae, radio spectral indices are usually less than around $0.3$ \citep{b31}.

Finally, for a completeness of this summary, it should be noted that among the Galactic and some Large Magellanic Cloud 
(LMC) SNRs, existence of significant spectral break and steepening at higher radio frequencies were also observed 
(Crawford et al.\@ 2008; Xiao et al.\@ 2008; Bozzetto et al.\@ 2010; de Horta et al.\@ 2012). The best example is 
the radio continuum spectrum of Galactic SNR G180.0-1.7 (S147) for which the spectral break is around 1.5 GHz, 
with low-frequency spectral index around 0.3 and high-frequency spectral index around 1.20 (Xiao et al.\@ 2008). 
The compression of the local magnetic field and shift of the turn-over in the Galactic radio spectrum to higher 
frequencies is one of the possible explanation of this kind of spectrum. On the other hand, synchrotron losses 
during the early phase of the SNR would cause a bend at a rather high frequency, but subsequent expansion of the 
SNR would shift the frequency toward about 1.5 GHz (Xiao et al.\@ 2008). Some of the models that try to explain 
"concave down" spectra of some Galactic and extragalactic objects (including synchrotron losses as well as finite 
emission region) are presented in the following papers: Bregman (1985), Biermann \& Strittmatter (1987), 
Heavens \& Meiseheimer (1987).

It is very difficult to derive accurate integrated flux densities. The small number of data points 
(with associated errors) and the dispersion of flux densities at the same frequencies pose a practical problem 
in determining the (integrated) radio spectral index and, generally, the shape of radio continuum. In that sense, it 
should be noted that present knowledge of radio spectral indices as well as of the overall shape of the Galactic SNR 
radio continuum spectra is far from being precise (at least for the majority of Galactic SNRs). New observations 
usually lead to non-negligible changes in the measured spectral indices. For example, although for some mixed-morphology 
SNRs Greens's catalog \citep{b7} gives rather steep spectral index, new observations given in \citet{b33} and \citet{b34} 
give slightly flatter, but still steep ($\alpha>0.5$) spectra (e.g.\@ for G53.6-2.2, $\alpha=0.75\rightarrow0.50$; 
G93.7-0.2, $\alpha=0.65\rightarrow0.52$; G116.9+0.2, $\alpha=0.61\rightarrow0.57$; G160.9+2.6, $\alpha=0.64\rightarrow0.59$). 
In this paper are considered only those SNRs for which the flux densities are on the scale of \citet{b1} and have known 
errors $<20\%$. SNRs with only four or less data points are not discussed in this paper.

Table 1 shows the list of several Galactic SNRs, with well defined radio spectra in the sense of previously mentioned 
criteria, and their basic properties: type, integrated radio spectral index $\alpha$, presence of low-frequency 
turn-over in overall spectrum (t.o.), distance $d$, mean diameter $D$, approximate age $t$, association/interaction 
with molecular cloud (MC), and detection in $\gamma$-rays ($\gamma$).

The question of nature of the flat spectral indices in some Galactic SNRs is, in fact, still open. In Sections 2 -- 4 a 
brief revision of proposed explanations (with the strongest arguments in their favor) is presented. 
Section 4 also deals with high energy properties of the SNRs with flat radio spectra focusing on their detection in 
$\gamma$-rays with the Large Area Telescope onboard Fermi $\gamma$-ray Space Telescope ({\it Fermi} LAT). In Section 5, model 
which includes significant intrinsic thermal bremsstrahlung radiation is discussed. Finally, Section 6 concludes the paper.

\section[Models involving Fermi II mechanism]{Models involving second-order Fermi mechanism}

Although Fermi I (DSA) is generally believed to be a primary mechanism for particle acceleration in SNRs, 
Fermi II (stochastic acceleration, see Liu et al.\@ 2008 for details) process in the shock turbulent vicinity could also 
be significant in some cases.

\citet{b20} used results of Dr\"{o}ge et al.\@ (1987) who solved the steady-state transport equation for the 
volume-integrated phase space distribution function of relativistic electrons in the vicinity of a plane-parallel 
shock wave allowing simultaneously for energy gain by multiple shock crossing and by second-order Fermi momentum 
diffusion upstream and downstream of the shock. They concluded that the observed dispersion in spectral index values 
below $\alpha=0.5$ is attributed to a distribution of low upstream plasma $\beta$ values for different SNRs.

Plasma $\beta$ is defined as follows: \begin{equation}
\beta=\frac{P}{P_{\rm{mag}}}=\frac{8\pi P}{B^{2}}=\frac{2}{\gamma}\left(\frac{V_{\mathrm{sound}}}{V_{\rm A}}\right)^{2}=\frac{8\pi nk(T_{\rm e}+T_{\rm i})}{B^{2}},
\end{equation}
where $P$ is the gas pressure, $P_{\rm{mag}}$ the magnetic pressure, $B$ the magnetic induction, $n$ the number density, 
$k$ the Boltzmann constant, $T_{\rm e}$ and $T_{\rm i}$ the electron and ion temperatures (if different, or just temperature $T$ 
of the medium), $V_{\mathrm{sound}}$ the adiabatic sound speed, $V_{\rm A}$ the Alfv$\mathrm{\acute{e}}$n speed and $\gamma$ 
the adiabatic index. For example, temperatures of around $10^{4}$ K ($\sim$ 1 eV) and number densities of around $0.1\ \rm{cm^{-3}}$ 
lead to the plasma $\beta$ of 0.16 for the average value of Galactic magnetic field (magnetic induction) of around 
$5\ \rm{\mu\rm{G}}$. Higher densities (as well as higher temperatures) would rise the plasma $\beta$. Also, higher magnetic 
fields will result in lower values of plasma $\beta$ for the same set of temperatures and densities.

\citet{b20} also pointed out that values of $\beta\simeq0.05$ are sufficient to yield $\alpha\simeq0.2$ for shock 
compression ratios higher than (or equal) 2.5 (assuming, also, constant spatial diffusion coefficient; see equations 
6 and 7 in Schlickeiser \& F\"{u}rst 1989).

On the other hand, in their work, they did not account for the possibility of compression ratios higher than 4 that could be 
expected in some (evolutionary) older SNRs (spectral index becomes 0.5 for compression ratio of 4 independently of 
plasma $\beta$). Dr\"{o}ge et al.\@ (1987) emphasized that the flat spectra could be obtained for compression 
ratios that tend to the value of 4 if the spatial diffusion coefficient increases with momentum and $\beta=1$. However, 
in the case of $\beta=1$ plasma and constant spatial diffusion coefficient, flat spectral indices are obtained for 
compression ratios higher than 4 (which is not discussed in Dr\"{o}ge et al.\@ 1987). Adiabatic index is assumed 
to be $5/3$ in all the relevant equations.

\citet{b13} discussed the particle acceleration process at a (parallel) shock wave, in the presence of the second-order Fermi 
acceleration in the turbulent medium near the shock, as an alternative explanation for the observed flat radio spectra 
of some Galactic SNRs in molecular clouds. He found that even very weak shocks may produce very flat cosmic-ray particle 
spectra in the presence of momentum diffusion (using power-law form for the spatial diffusion coefficient -- $\kappa\propto p^{0.67}$, 
assuming shock compression ratio of 4 and applying an effective jump condition for the Alfv$\mathrm{\acute{e}}$n velocity 
-- $V_{{\rm A},1}=V_{{\rm A},2}$, where indices 1 and 2 refer to the upstream and downstream of the shock, respectively). 
Ostrowski (1999) pointed out that as a consequence, the dependence of the particle spectral index of the shock 
compression ratio can be weaker than that predicted for the case of pure first-order acceleration. His analysis 
was based on the model of Ostrowski \& Schlickeiser (1993) that improved the work of Dr\"{o}ge et al.\@ (1987) and 
\citet{b20} deriving a simplified kinetic equation in the momentum space valid for the case of small momentum diffusion 
and also allowing the description of stationary spectra at the shock.

\citet{b13} emphasized that shock waves with Alfv$\mathrm{\acute{e}}$n speed non-negligible in comparison to the shock 
velocity are responsible for generating the flat particle distribution. He noted that for the ratio of 
Alfv$\mathrm{\acute{e}}$n speed and shock velocity (inverse of the so called Alfv$\mathrm{\acute{e}}$n Mach number, 
see equation 4) around 0.1 the second-order Fermi process can substantially modify the energy spectrum of the 
particles accelerated at the shock and that the model holds for values of that ratio less than 0.2.

Previously mentioned Alfv$\mathrm{\acute{e}}$n Mach number of a shock is determined by \citep{b2}:
\begin{equation}
M_{\rm A}=\frac{V_{\rm s}}{V_{\rm A}}\approx460\ V_{{\rm s}8}\ n_{\rm a}^{1/2}\ /\ B_{-6}, 
\end{equation}
where $V_{\rm s}$ is the shock velocity, $V_{\rm A}$ is the Alfv$\mathrm{\acute{e}}$n 
velocity and $V_{{\rm s}8}$ is the shock velocity in $10^{8}\ \rm{cm}\ \rm{s}^{-1}$, $n_{\rm a}$ is the ionized 
ambient gas number density in $\rm{cm}^{-3}$ and $B_{-6}$ is the local magnetic induction, just before the shock, 
measured in $\mu\rm{G}$. Alfv$\mathrm{\acute{e}}$n Mach number is generally an important parameter as the ratio 
of acceleration rates of the Fermi I (DSA) and Fermi II (second-order) processes scales as $M_{\rm A}^{-2}$ \citep{b16}. 
As the bulk post-shock flow is expected to be super-Alfv$\mathrm{\acute{e}}$nic it is logical to conclude that the DSA 
is more rapid process.

The corresponding analysis was successfully applied to the SNRs W44 and IC443. Although \citet{b13} could not reproduce the 
steep spectral index (0.55) in the case of SNR 3C391, using the spectral index value of 0.49 \citep{b7}, this model could be 
justified (for shock velocity of around 300 km/s and inter-clump density of around $10\ \rm{cm^{-3}}$, as in Ostrowski 1999, 
this model could account for observable spectral index of 0.49 if magnetic induction is $\sim 17\ \rm{\mu\rm{G}}$; see 
equations 2.6-2.9 from Ostrowski 1999). It is worth mentioning that compression ratios higher than 4 (not discussed in 
Ostrowski 1999) would give, in this model, lower magnetic induction estimates.

It is worth noting that SNRs expanding in molecular clouds mostly evolve in the clumpy medium. The dominant part of 
the cloud mass is contained in the compact dense clumps filling only around 10\% or less of the cloud volume and the rest 
of the cloud is filled with inter-clump gas of average number density of $10\ \rm{cm^{-3}}$ (Chevalier 1999; Ostrowski 1999). 
However, one must bear in mind that SNR precursors could contribute to modification of physical conditions in the immediate 
pre-shock ISM (Arthur 2012).

The fact that many SNRs with flat spectral indices are also connected with molecular cloud environment could be a sign that 
models like \citet{b13} are more appropriate in these cases. As the radiative shock has been revealed by the {\it Spitzer} 
IRS slit observation of SNR 3C396, that expands through the clumpy environment, model of Ostrowski (1999) could possibly 
account for the observed flat radio spectral index (Hewitt et al.\@ 2009; Su et al.\@ 2011). For inter-clump density of 
around $1\ \rm{cm^{-3}}$ and for shock velocities of 120 km/s and 350 km/s (Hewitt et al.\@ 2009), magnetic induction of 
$10-29\ \rm{\mu\rm{G}}$ is needed, respectively, to account for observable radio spectral index of 0.34 (see equations 
2.6-2.9 from Ostrowski 1999). For a blast wave velocity of 870 km/s (Su et al.\@ 2011), needed value of magnetic induction 
is higher -- around $72.5\ \mu\rm{G}$. In the case of SNR W28, for shock velocity of around 80 km/s and inter-clump density 
of around $5\ \rm{cm^{-3}}$ (Rho \& Borkowski 2002), model of Ostrowski (1999) could account for observable spectral index 
of 0.35 if magnetic induction is $\sim 14\ \rm{\mu\rm{G}}$. For higher inter-clump density of $10\ \rm{cm^{-3}}$, needed 
magnetic induction would be around $20\ \rm{\mu\rm{G}}$. Finally, HB21 is known to be interacting with clumpy molecular 
clouds (Pannuti et al.~2010 and references therein). If we assume, approximately, 130 km/s for the velocity of radiative 
shell and the mean ambient \mbox{H\,{\sc i}} density of $3.7\ \rm{cm^{-3}}$ (Koo et al.~2001 and references therein), model 
of Ostrowski (1999) could account for observable radio spectral index of 0.36 if magnetic induction is 
$\sim 19\ \rm{\mu\rm{G}}$. In all of these cases compression ratio of 4 was assumed.

Finally, it must be emphasized that there are several SNRs, with flat spectral indices, that are generally associated 
with relatively lower densities such as G82.2+5.3 (W63), with $\alpha=0.44$ (Higgs et al.~1991; Mavromatakis 2004; 
Gao et al.\@ 2011), and G166.0+4.3 (VRO 42.05.01), with $\alpha=0.33$ (Guo \& Burrows 1997; Leahy \& Tian 2005; 
Bocchino et al.\@ 2009; Gao et al.\@ 2011). Both SNRs are classified as mixed-morphology ones. W63 is probably 
expanding in the wind blown bubble and VRO 42.05.01 has probably broken out of the cloud within which it formed, 
expanded across an interstellar tunnel or cavity, and is now interacting with the material that forms the 
opposite tunnel wall (Pineault et al.\@ 1987; Burrows \& Guo 1994; Guo \& Burrows 1997; Mavromatakis 2004). 
If the model of \citet{b20} is acceptable for these SNRs, then the observed spectral indices lead to 
the plasma $\beta$ around 0.17 and 0.05 for W63 and G166.0+4.3, respectively, in the case of compression ratio of 3.9 
and constant spatial diffusion coefficient. Slightly higher values of plasma $\beta$ are obtained for compression ratio 
of 2.5. If power-law index of the spatial diffusion coefficient dependence of momentum equals 0.67, then observed spectral 
indices could not be obtained by the model of Dr\"{o}ge et al.\@ (1987). Finally, if plasma $\beta=1$ than compression ratios 
of around 6.95 and 12.3 correspond to the observed spectral indices of W63 and G166.0+4.3, respectively (for constant spatial 
diffusion coefficient). These estimates will, on the other hand, lead to (unrealistically) low isothermal Mach number, 
fully radiative SNR shocks.

\section[Model involving high compression ratios]{Models involving high compression ratios}

The flatter spectra may also be due to compression ratios greater than four at fully radiative (isothermal) shocks 
(in the post Sedov-Taylor phases of SNR evolution). In the case of parallel isothermal shock, compression ratio 
$\chi$ is equal to a square of isothermal Mach number ($M_{\rm T}$) so that test particle DSA gives:
\begin{equation}
\alpha=\frac{3}{2(\chi-1)},\ \ \ \ \chi=M_{\rm T}^{2}.
\end{equation}
In Figure 1 it is shown that this explanation, for observable flat spectral index range, $\alpha\in(0.25-0.5)$, 
holds only for very low isothermal Mach number fully radiative shocks (isothermal Mach numbers around 2--3) 
in the test-particle DSA.

Of course, one must bear in mind that the onset of the so called momentum-conserving snowplow phase occur 
only near the end of a SNR lifetime, or not occur at all and even a pressure-driven snowplow model is just 
the asymptotic case \citep{b29}. Of course, local regions of the SNR blast wave may leave the Sedov-Taylor phase earlier 
at those places where the ambient density is much higher than average, though the bulk evolution still corresponds to 
energy conserving phase \citep{b15}. This means that different parts of an SNR may be in different phases so that 
different shock velocities and compression ratios within the SNR are possible, which could have repercussions 
on the integrated radio spectrum.

\begin{figure}[h!]
\includegraphics[width=84mm]{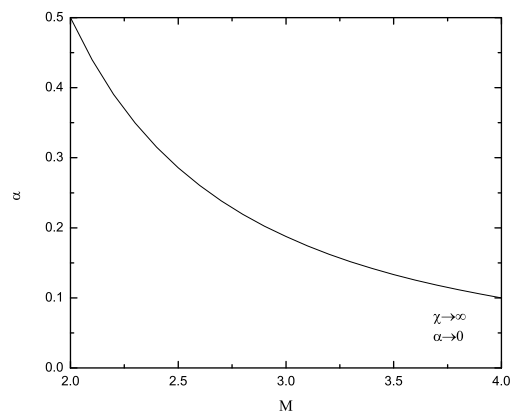}
 \caption{The change of integrated radio spectral index with isothermal Mach numbers of fully radiative (isothermal) 
SNR shock wave.}
\end{figure}

It must be noted that the values of adiabatic index slightly less than $5/3$ (due to high temperatures behind the shock front 
and/or because the molecules in the cloud, with which SNR interacts, will cool more efficiently than the ambient medium or 
the ejecta as discussed in Davis 2007) would reproduce some of the observable flat spectra in the case of parallel adiabatic 
shock wave and test-particle DSA (see equation 2). On the other hand, such values are generally not expected in the 
post-shock medium. Finally, Anand (2012) derived and discussed the jump relations across a shock in non-ideal gas flow. 
It is interesting to note that even for very small non-idealness of gas, using equation (5) from Anand (2012), for the 
compression ratio, flat spectral indices are obtained for the large range of shock velocities in the framework of DSA. 
Of course, applicability of the non-ideal gas equation of state in the case of the particular ISM plasmas is questionable.

\section{High energy properties of the SNRs with flat integrated radio spectral indices}

Since there is a significant connection between the majority of Galactic SNRs with flat integrated radio spectrum and their 
detection in $\gamma$-rays, the high energy properties of these SNRs are now discussed (see Table 1). Apart from historical 
(e.g.\@ Tycho and Cassiopeia A -- Abdo et al.~2010e; Giordano et al.~2012) and young TeV-bright SNRs 
(e.g.\@ RX J1713.7-3946 and Vela Jr. -- Abdo et al.~2011; Tanaka et al.~2011), among the SNRs recently 
detected with {\it Fermi} LAT, there is a huge class of SNRs that interact with molecular clouds as well as of the 
evolved SNRs without molecular cloud interaction (e.g.\@ Cygnus Loop, S147 -- Katagiri et al.~2011; Katsuta et al.~2012).

$\gamma$-ray emission from SNRs can be produced by few different, non-exclusive, mechanisms. Electrons and positrons in the 
SNR shell can interact by inverse Compton scattering with ambient photon fields (such as cosmic microwave background and 
infrared radiation) to produce $\gamma$-rays (Helder et al.\@ 2012). Also, electron-atom or electron-molecule interactions 
in a dense medium result in bremsstrahlung radiation (Reichardt et al.~2012). Mentioned mechanisms involve electrons 
accelerated up to the sufficient energies to produce the observed $\gamma$-rays (leptonic scenario). $\gamma$-rays 
can also be produced by neutral pion decay, where neutral pions result from the collision of accelerated protons 
(or heavier nuclei) with nucleons of the ambient gas (hadronic scenario).

In the case of SNRs that interact with molecular clouds, hadronic scenario is probable and one of the spectral signatures 
is steepening in the GeV band. Generally, absence of non-thermal X-ray emission favors hadronic origin of the observed 
$\gamma$-ray emission. The rapid steepening of the spectrum above few GeV is the kind of signature expected from the 
re-acceleration of pre-existing cosmic-rays, where high energy cosmic-rays escape from the SNR confinement region 
(Reichardt et al.~2012 and references therein). Uchiyama et al.\@ (2012) emphasized that it has not been clear yet 
whether the $\gamma$-ray emission is produced by escaping or by trapped cosmic-rays, since shocked molecular clouds 
inside an SNR could also be the sites of efficient $\gamma$-ray production. Although the majority of SNRs that interact 
with molecular clouds and are detected in $\gamma$-rays are those with flat radio spectra (e.g.\@ W44, IC443, W28, W49B, 
3C391, W51C, CTB37A, HB21 -- Abdo et al.~2009; Abdo et al.\@ 2010abcd; Castro \& Slane 2010; Reichardt et al.~2012), 
it must be mentioned that several SNRs that interact with molecular clouds and have spectral indices of 0.5 are also 
detected in $\gamma$-rays with {\it Fermi} LAT (e.g.\@ G8.7-0.1, G109.1-1.0, G349.7+0.2 -- Castro \& Slane 2010; 
Ajello et al.~2012; Castro et al.\@ 2012).

Reichardt et al.~(2012) emphasized that SNR HB21 belongs to the group of low-luminosity, GeV-emitting SNRs 
(such as Cygnus Loop or S147), which are clearly less luminous than the first GeV-emitting SNRs that were 
discovered (such as W51C, IC443, W49B). The spectral break is found at lower energy in HB21 than in the 
case of the luminous SNRs. Reichardt et al.~(2012) suggested that the $\gamma$-ray emission from HB21 can
be understood as a combination of emission from shocked/illuminated molecular clouds, one of them coincident 
with the SNR shell itself.

SNR G260.4-3.4 (Puppis A), with radio spectral index of 0.5 (Green 2009), was also observed at $\gamma$-rays 
(Hewitt et al.~2012) and it represents an interesting transitional case between young SNRs still evolving into 
a circumstellar medium (e.g.\@ Cas A), and older SNRs which are interacting with large, dense molecular clouds 
(e.g.\@ IC 443).

In the case of Kes 67, not yet detected by {\it Fermi} LAT, the bright pulsar wind nebula HESS J1825-137 is extended near 
the position of the SNR (Grondin et al.~2011) thus making the analysis highly complicated. Similarly, near the position of 
SNR 3C396, a radio-faint $\gamma$-ray pulsar (PSR J1907+0602) powering a bright TeV pulsar wind nebula (MGRO J1908+06) is 
located, substantially complicating the analysis (Abdo et al.~2010f).

Possible interaction with molecular cloud in the case of SNR 3C434.1 is indicated by Jeong et al.~(2012), which makes 
this remnant an interesting target for analysis of possible $\gamma$-radiation in its direction. On the other hand, 
there are no point sources included in the LAT 2-year point source catalog (Nolan et al.~2012) closer than 2 degrees 
to the position of SNR 3C434.1. Also, there are no extended sources closer than 10 degrees to the position of this remnant. 
A detailed study of the $\gamma$-emission in the direction of SNR 3C434.1 is planned for a work in the near future. 

\subsection[Model of Uchyama et al.]{Model of Uchyama et al.}

\citet{b24} analyzed $\gamma$-ray emission from SNRs interacting with molecular clouds. They showed that the simple 
re-acceleration of pre-existing cosmic-rays by the process of diffusive shock acceleration at a cloud shock is generally 
sufficient to power the observed $\gamma$-ray emission through the decays of neutral pions. Neutral pions are produced in 
hadronic interactions between high-energy protons (nuclei) and gas in the compressed-cloud layer. \citet{b24} proposed 
that the radio emission may be additionally enhanced by the presence of secondary electrons/positrons, i.e.\@ the products 
left over from the decay of charged pions, created due to cosmic-ray nuclei colliding with the background plasma. They 
concluded that presence of secondary electrons/positrons may also explain the flat spectral radio indices of some 
mixed-morphology SNRs.

\citet{b24} applied their analysis to SNRs W51C, W44 and IC443. Of course, this kind of explanation holds only for those SNRs 
that interact with molecular clouds assuming further that hadron scenario for $\gamma$-radiation is significant. Another 
restriction of this model is the fact that \citet{b24} used relations that are applicable in the case of the Sedov-Taylor 
phase which are not suitable for mixed-morphology SNRs. On the other hand, the simple re-acceleration of pre-existing cosmic-rays 
and subsequent compression alone would not fully explain the $\gamma$-rays associated with cloud-interacting SNRs. It must be 
noted that GeV and TeV $\gamma$-rays outside the southern boundary of SNR W28 may be explained by molecular clouds 
illuminated by runaway cosmic-rays (Uchiyama et al.~2010 and references therein). \citet{b24} also assumed pre-existing 
cosmic-rays in the cloud to have the same spectra as the Galactic cosmic-rays in the vicinity of the Solar system although 
the ambient cosmic-rays in the pre-shock cloud may deviate from the Galactic pool due to the runaway cosmic-rays that have 
escaped from SNR shocks at earlier epochs.

Another model by \citet{b3} was also proposed assuming the possibility of electron acceleration by 
magnetohydrodynamical turbulence behind high density shocks (Vink 2012). In their model, the calculated radio 
spectrum fits that observed from the SNR IC443 shell because of the large shock compression ratio and to a lesser 
extent because of the effect of the second-order Fermi acceleration. It is worth mentioning that this model involves 
a lepton scenario of $\gamma$-ray emission from SNRs interacting with molecular clouds. 

\section[Model involving thermal radiation]{Model involving significant intrinsic thermal bremsstrahlung radiation}

Recently, it was shown that the observations over a very broad range of radio frequencies reveal a curvature in the spectra 
of some evolutionary older Galactic SNRs expanding in high density environment (Tian \& Leahy 2005; 
Uro{\v s}evi{\' c} \& Pannuti 2005; Leahy \& Tian 2006; Uro{\v s}evi{\' c}, Pannuti \& Leahy 2007; 
Oni\'c \& Uro{\v s}evi{\'c} 2008). The NDSA effects could not be responsible for a characteristic shape of 
the integrated radio spectra in these cases. The presence of intrinsic thermal bremsstrahlung radiation was proposed 
as an explanation of the "concave up" radio spectrum (Uro{\v s}evi{\' c} 2000; Uro{\v s}evi{\' c} et al.\@ 2003ab; 
Oni\'c et al.\@ 2012 and references therein). Presence of the cooled thermal X-ray electrons in post Sedov-Taylor SNRs, 
higher ambient densities, interaction with or expansion in molecular cloud are main arguments used by the model that 
includes significant intrinsic thermal bremsstrahlung emission (thermal model). Detections of the low-frequency turn-overs, 
in the integrated radio spectrum, related with thermal absorption linked to SNRs (Brogan et al.\@ 2005; 
Castelletti et al.\@ 2011) are important in justification of this model as one can then predict 
the significance of thermal emission at higher frequencies. Generally, linear polarization measurements 
(that give the upper limits for the thermal component of the radio emission), possible detections in 
$\mathrm{H}\alpha$ as well as the observations of radio recombination lines associated with SNRs should accompany 
this kind of analysis.

It must be emphasized that most of the potential targets for testing the thermal model hypothesis fall 
in the mixed-morphology class of SNRs. Because of the fact that these SNRs appear to be in the radiative 
phases of their evolution, with shock velocities less than 200 km/s (Vink 2012), the ensemble of cooled 
thermal X-ray electrons could exist, supporting the thermal model \citep{b11}. In addition, recent discovery of 
strong radiative recombination continua (RRC), seen in the X-rays of several SNRs of this type, represents 
definite evidence that their plasma is recombining (Yamaguchi et al.\@ 2012). RRCs in their X-ray spectra are 
indicative of higher densities in interiors of these remnants (Vink 2012). Thus, mixed-morphology SNRs are 
characterized by a more/less uniform high interior density with medium hot temperatures, rather than very high 
temperatures and very low interior densities, as in the case of typical shell remnants. Also, among the SNRs with detected 
RRC, the most of them (if not all) have flat radio spectral index (Sawada \& Koyama 2012; Uchida et al.~2012; 
Vink 2012). Finally, as it was emphasized earlier, the morphology of these SNRs is difficult to explain with standard 
SNR evolutionary phases \citep{b26} and several models were proposed in the current literature (White \& Long 1991; 
Cox et al.\@ 1999). An X-ray emitting overionized plasma is the result of an early heating followed by a rapid cooling, 
but the mechanism responsible for such processes in mixed-morphology SNRs is still not completely understood (Miceli 2011).

\citet{b11} have recently analyzed the integrated radio spectra of 3 SNRs (IC443, 3C391, 3C396) for which the thermal model 
could be a natural explanation. All of the 3 SNRs are characterized by flat spectral indices (see Table 1) based on a simple 
power-law (synchrotron) fit and all of them interact with molecular clouds. For IC443 and 3C391 the low-frequency turn-over 
is due to thermal absorption linked to the SNRs and for SNR 3C396 the nature of the observed low-frequency turn-over is not 
yet clear. It must be emphasized that most of the SNRs that are the best targets for testing thermal model are also 
characterized by the flat spectral indices. The observed flat spectral indices could be just the apparent manifestation of 
the presence of thermal component. It must be noted that there are no predictions of curved -- "concave up" integrated radio 
spectra in the cases of the explanations presented in Sections 2--4 for the flat radio spectra seen in some SNRs.

The shock wave dissociates molecules, or raises the gas temperature so that previously inaccessible degrees of freedom 
become accessible. The compression ratio for the parallel shock wave, including the possibility of jump in adiabatic index, 
is as follows: \begin{eqnarray}
&\chi=&\frac{\gamma_{2}\left(M^{2}+\frac{1}{\gamma_{1}}\right)}{(\gamma_{2}-1)\left(M^{2}+\frac{2}{\gamma_{1}-1}\right)}+\nonumber\\
&&\quad\quad\quad\quad+\frac{\sqrt{M^{4}+M^{2}\frac{2(\gamma_{1}-\gamma_{2}^{2})}{\gamma_{1}(\gamma_{1}-1)}+\left(\frac{\gamma_{2}}{\gamma_{1}}\right)^{2}}}{(\gamma_{2}-1)\left(M^{2}+\frac{2}{\gamma_{1}-1}\right)},
\end{eqnarray}
where $\gamma_{1}$ and $\gamma_{2}$ are adiabatic indices in upstream and downstream region, respectively. If 
$\gamma_{1}=\frac{7}{5}$ (diatomic molecules like $\rm{H_{2}}$) and $\gamma_{2}=\frac{5}{3}$ (fully ionized gas) 
it is easily seen that (upstream) adiabatic Mach number $M=20$ leads to $\alpha\approx0.51$ so that for weaker shocks the steeper 
spectra will be preferable. As the ambient gas is usually pre-ionized by radiative precursor, the similar conclusion (for slightly 
smaller adiabatic Mach numbers) could be drawn from the simple relation for constant adiabatic index of $5/3$ (described in 
Section 1 using equation 2, see Figure 2). This means that we expect that evolutionary old SNRs, with velocities 
smaller than around $200\ \rm{km\ s^{-1}}$, have steeper synchrotron spectral indices in the framework of 
test-particle DSA. This can partially account (except high parameter uncertainties) for steeper synchrotron spectral 
indices determined from the thermal plus non-thermal model fits \citep{b11}.
\begin{figure}[h!]
\includegraphics[width=84mm]{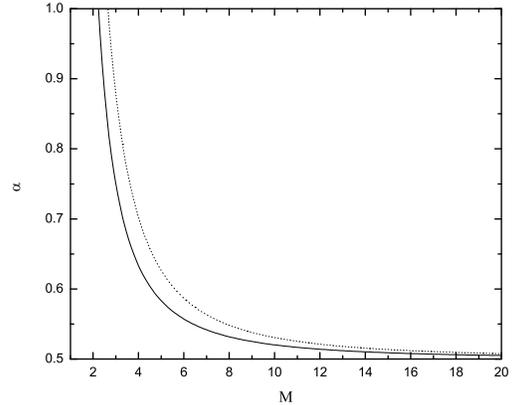}
 \caption{Change of the radio spectral index with Mach number for slow shocks using equations (2) -- full line and (6) -- 
dotted line.}
\end{figure}

Radio spectral index variations, both with frequency and location within several Galactic SNRs were detected 
(F\"{u}rst \& Reich 1988; Anderson \& Rudnic 1993; Zhang et al.\@ 1997; Leahy \& Roger 1998; 
Leahy et al.\@ 1998; Tian \& Leahy 2005; Leahy 2006; Ladouceur \& Pineault 2008). In order to have spectral changes with 
position, physical conditions that alter the radio spectrum must change with position. Of course, spatial radio spectral 
index variations are naturally expected in the case of composite SNRs due to the presence of pulsar wind nebula. On the 
other hand, variations in $\alpha$ are detected in both shell and mixed-morphology SNRs. If the spectral index is changing 
with frequency, generally, there is a possibility that two different radiation mechanisms are producing the radio continuum 
emission. Integrated radio spectrum with "concave up" curvature in the case of older SNRs, where efficient NDSA process is 
not expected, can be produced by adding two different emission spectra along the same line of sight or within the same 
emitting volume (Leahy \& Roger 1998). These kind of spectra could be due to different electron populations or because 
of the significant intrinsic thermal bremsstrahlung radiation \citep{b11}. Linear polarization measurements are of the great 
importance as spectral flattening associated with high density regions of an SNR which do not show high linear polarization 
could be naturally explained by the thermal model.

It is worth mentioning that there are some other processes that could shape the integrated radio continuum at higher 
frequencies (between 10-100 GHz) in several Galactic SNRs, such as the spinning dust emission (Draine \& Lazarian 1998; 
Scaife at al.\@ 2007). The spinning dust model will be responsible for a creation of a "hump" in the high-frequency 
radio spectrum (see Figure 5 from Scaife at al.\@ 2007). Of course, as it was emphasized it the Section 1, present knowledge 
on integrated radio spectrum is generally unsatisfactory for firm quantitative analysis of the proposed models so as to 
the contribution of dust emission. More data at radio frequencies higher than 1 GHz, e.g.\@ those from ATCA (The Australia 
Telescope Compact Array), are expected to shed a light on the existence of the so called "radio thermally active" SNRs as 
well as on the issue of the flat spectral indices. It must be mentioned that there is a general problem in observing at 
the higher radio continuum frequencies ($>$30 GHz) due to atmospheric absorption issues. Ground-based radio astronomy 
is limited to high altitude sites such as in the case of ALMA (Atacama Large Millimeter Array).

\subsection{The results regarding thermal model for Galactic SNRs}

Except the supernova remnants G31.9+0.0 (3C391), G39.2-0.3 (3C396), G189.1+3.0 (IC443), analyzed in \citet{b11}, that 
represent the best targets for justification of the thermal model, in this paper, the integrated radio spectra of 
SNRs G6.4-0.1 (W28), G18.8+0.3 (Kes 67), G34.7-0.4 (W44), G43.3-0.2 (W49B), G89.0+4.7 (HB21), G94.0+1.0 (3C434.1) were 
also discussed. All of these SNRs are associated with dense environment (see Table 1). Integrated flux densities are 
collected from the literature keeping in mind the criteria defined in Section 1. In the case of presence of spectral 
turn-over at low frequencies only data which are not influenced by the possible thermal absorption (or maybe by 
synchrotron self-absorption) were used in thermal model fits \citep{b11}. Simple sum of non-thermal and thermal components, 
represented by corresponding power-laws, (see equation 18 and corresponding discussion in Oni\'c et al.\@ 2012) is assumed. 
For frequencies in $\mathrm{GHz}$, the relation for the integrated flux density can be written as follows: \begin{equation}
S_{\nu}=S_{1\ \!\!\mathrm{GHz}}^{\mathrm{NT}}\ \left(\nu^{-\alpha}+\frac{S_{1\
\!\!\mathrm{GHz}}^{\mathrm{T}}}{S_{1\
\!\!\mathrm{GHz}}^{\mathrm{NT}}}\ \nu^{-0.1}\right)\ [\mathrm{Jy}],
\end{equation} where: $S_{1\ \!\!\mathrm{GHz}}^{\mathrm{T}}$ and $S_{1\ \!\!\mathrm{GHz}}^{\mathrm{NT}}$ are flux densities 
corresponding to thermal and non-thermal components, respectively. Weighted least squares fit 
(influenced by the instrumental errors) were applied. In the case of all of these six SNRs thermal model fit 
is either not significant (contribution of the thermal component at 1 GHz is near zero) or the fit quality 
is not satisfactory (and associated parameter errors exceed values of the corespondent parameters). Pure 
non-thermal fit parameters for these SNRs correspond to ones given in the present literature (see Table 1).

In the case of SNR W28, low-frequency spectral turn-over in the integrated spectrum is possible as flux density at 30.9 MHz 
has a rather low value in the sense of scatter from the power-law fit (see Figure 5 in Dubner et al.\@ 2000). The associated 
error for that flux density is 20\% (Kassim 1989). On the other hand, there is a great spread in flux density values, 
especially at 2.7 GHz, which prevents firm conclusions. The great spread in the integrated flux densities at the same frequencies 
is very good seen in the case of SNRs W49B and W44 (Morsi \& Reich 1987; Kassim 1989; Taylor et al.~1992; Kovalenko et al.~1994; 
Lacey et al.~2001; Castelletti et al.\@ 2007; Sun et al.~2011). For SNR W44 there is no apparent low-frequency 
spectral turn-over in the integrated radio spectrum, only the localized one, towards the southeast border of the SNR, 
likely due to the free-free absorption from ionized gas in the post-shock region at the SNR/molecular cloud interface 
(Castelletti et al.\@ 2007). Detected turn-over at lower frequencies for W49B is likely due to the extrinsic free-free absorption 
by an intervening cloud of thermal electrons (Lacey et al.~2001). There is no apparent low-frequency spectral turn-over in 
integrated spectra of SNRs Kes 67, HB21 and 3C434.1 (Reich et al.~1983; Goss et al.~1984; Landecker et al.~1985; Kassim 1989; 
Milne et al.~1989; Tatematsu et al.~1990; Kovalenko et al.~1994; Dubner et al.~1996; Zhang et al.~2001; Reich et al.~2003; 
Foster 2005; Kothes et al.~2006; Gao et al.~2011; Sun et al.~2011).

In the case of other SNRs that are listed in \citet{b7} as those with flat spectral indices (not plerions), expanding in 
high density environment, there are not enough data (under criteria presented in Section 1) for a detailed discussion. 
Generally, there is a significant number of SNRs listed in \citet{b7} for which the spectral indices are determined 
only by two or three data points. Some of the SNRs that could be interesting targets for analysis, in future, when more 
reliable data at different frequencies are obtained, are G49.2-0.7 (W51C) and G290.1-0.8 (MSH 11-61A).

A few words of caution regarding the justification of thermal model. SNR's expansion in complex environment and possible 
association with \mbox{H\,{\sc ii}} regions (and so possible thermal contamination) makes the study more complicated. The 
best example is the case of steep radio spectrum SNR G132.7+1.3 (HB3) for which the intrinsic thermal radiation was 
dismissed due to the actual overlap between the SNR and corespondent \mbox{H\,{\sc ii}} regions (Shi et al.~2008). 
Of course, the question on the presence of significant thermal radio emission from SNR HB3, in fact, still remains open as 
the possible \mbox{H\,{\sc ii}} region radiation contamination can not rule out the possibility of thermal bremsstrahlung 
emission from the SNR itself, it can just mask it.

Final conclusion is that our present knowledge on integrated flux densities prevents us from possibility of strong 
justification of the thermal model and limits the discussion to just a few Galactic SNRs analyzed in \citet{b11}. 

\section{Conclusions}

A considerable fraction of Galactic SNRs are characterized by the flat spectral indices ($\alpha<0.5$). In this paper, 
known models are summarized and discussed:

\begin{enumerate}
  \item There are several explanations of the observable flat radio spectra of some SNRs. Most of the models involve a 
significant contribution of second-order Fermi mechanism, but some of them discuss higher shock compressions, contribution 
of secondary electrons/positrons left over from the decay of charged pions, as well as the possibility of thermal 
contamination.   
  \item In the case of expansion in high density environment, thermal bremsstrahlung could theoretically shape the spectrum 
of SNRs. Lack of more high quality data constraints the discussion only to several SNRs. On the other hand, this model can 
naturally account to observable curved -- "concave up" radio spectra of some Galactic SNRs. New observations are expected 
to create a clear picture about the high frequency range (1-100 GHz) of the radio continuum, as well as about the question 
of the significant contribution of intrinsic thermal bremsstrahlung radiation. With instruments like the 
JVLA\footnote{The Karl G.\@ Jansky Very Large Array of the National Radio Astronomy Observatory is a facility of the National 
Science Foundation operated under cooperative agreement by Associated Universities, Inc.} and ATCA, it should be far easier 
to measure the spectral indices in individual observations. 
  \item The fundamental question about the origin of flat spectral indices still remains open. The more realistic picture would involve 
the action of more than one of the mentioned processes, in which some of them could be more or less prominent depending on 
the particular setup.
  \item Since there is a significant connection between the majority of Galactic SNRs with flat integrated radio spectra 
and their detection in $\gamma$-rays, as well as detection of RRC in their X-ray spectra, the analysis of high energy 
properties of these SNRs will be potentially very important.
\end{enumerate}

\acknowledgments

This work is part of Project No.\@ 176005 "Emission nebulae: structure and evolution" supported by the Ministry of Education, 
Science, and Technological Development of the Republic of Serbia.

\end{document}